\documentclass[showpacs,reprint,amsmath,amssymb,prl,floatfix]{revtex4-1}

\usepackage{graphicx}
\usepackage{dcolumn}
\usepackage{bm}
\usepackage{amsmath}

\begin{document}

\title{Organic Molecules on Wide-Gap Insulators: Electronic Excitations} 
\author{Wei Chen}
\author{Christoph Tegenkamp}
\author{Herbert Pfn\"{u}r}
\email{pfnuer@fkp.uni-hannover.de}
\affiliation{Institut f\"{u}r Festk\"{o}rperphysik, Leibniz Universit\"{a}t Hannover, 30167 Hannover, Germany}

\date{\today}

\begin{abstract}
The electronic excitation of a conjugated molecule-insulator interface, as exemplified by the adsorption of benzoic acid and its phenolic derivative on NaCl(001) surface, is addressed by many-body Green's function methods.
By solving the two-particle Bethe-Salpeter equation on top of the $GW$ quasiparticle energies, it turns out that instead of the intramolecular $\pi-\pi^\ast$ transition of the adsorbate, the lowest singlet excited state of the adsorbate system is essentially assigned to the transition from the surface valence band maximum to the $\pi^\ast$ state of the molecule, as a charge transfer excitonic effect.
An accurate description of this lowest electronic excitations confined at the interface requires the knowledge of a full excitonic Hamiltonian due to the sizable electron-hole exchange interaction.

\end{abstract}
\maketitle

A revived interest in the organic molecule and wide-gap insulator interfaces has been witnessed in the last decade.
Their promising applications in molecular electronics \cite{Joachim2000} are benefited from the merit that the electronic characteristics of the adsorbed molecule are barely perturbed by the underlying insulating surfaces \cite{Repp2005}, which enables the use of wide-gap insulator as a supporting substrate for a wide variety of technical applications.
The organic molecule-insulator interface is also relevant to the separation process seen in mining industry, namely various mineral species, usually rocksalt (NaCl) and sylvite (KCl) can be distinguished electrostatically when mixing with certain acids, e.g. salicylic acids (SA).
It is speculated that in the combined molecule-insulator system, the presence of the highest occupied (HOMO) and the lowest unoccupied molecular orbitals (LUMO) can enhance the charge transfer by electronic excitations by several orders of magnitude with respect to the bare insulators, because the molecular energy gap is usually much smaller than the band gap of the wide-gap insulators \cite{Malaske1998,Tegenkamp2002}.
When the adsorbate covered insulating surfaces are brought into contact, those excited electrons, as mobile electrons, can hop in-between the surfaces until the thermal equilibrium is reached and the Fermi levels are aligned across the interface.
This gives rise to a contact voltage due to the charge transfer, which tentatively explains the origin of the electrostatic separation.

Previous Kohn-Sham density functional theory (KS-DFT) calculations \cite{Chen2009, Chen2010} have identified that the adsorption of benzoic acid (BA) and its phenolic derivatives (e.g. SA) on alkali chloride surfaces indeed yields a much smaller effective KS gap than that of the bare surface.
However, the KS gap, defined as the difference between the eigenvalues of the lowest unoccupied and highest occupied KS states, is \textit{not} a realistic measure of excitation properties.
One of the basic issues with KS gap is that it is always smaller than the fundamental gap by a derivative discontinuity of the exchange-correlation energy functional \cite{Perdew1982,*Perdew1983} even if the discontinuity is present in the functional \cite{Kummel2008}, and this error could be sizable for wide-gap insulators \cite{Gruning2006}.
This gap problem can be alleviated either by hybrid functionals incorporating nonlocal Fock exchange within the generalized KS scheme \cite{Seidl1996}, or ultimately from the quasiparticle energies in the $GW$ approximation of the self-energy \cite{Hedin1965}.
On the other hand, in context of the electronic excitations at the molecule-insulator interface, the electron-hole (\textit{e-h}) interaction is not directly accessible from the fundamental gap within a single-particle picture, but only available in time-dependent DFT \cite{Runge1984} and the two-particle many-body perturbation theory (MBPT), namely the Bethe-Salpeter equation (BSE) approach \cite{Salpeter1951}.
The excitonic effect has been found prominent for some organic molecules \cite{Rohlfing1999,Palummo2009,Gruning2009} and wide-gap insulator surfaces \cite{Rohlfing2003} in terms of MBPT.
Nevertheless, little is known about the electronic excitations when organic molecules are adsorbed on insulating surfaces.
In this Letter, we demonstrate that, based on $G_0W_0$ and BSE calculations of BA and SA adsorbed on NaCl(001) surface, the lowest singlet excited state for a $\pi$-conjugated molecule-insulator interface involves the contributions from both the surface and the adsorbate molecule, instead of the intramolecular $\pi-\pi^\ast$ transition.
This is not readily expected provided that the coupling of the molecule and the surface is weak.

The $G_0W_0$ and BSE calculations were performed on top of the KS-DFT eigenvalues and eigenstates within local density approximation (LDA).
The ground-state geometries were also optimized using DFT-LDA.
We employed plane-wave basis sets with a cutoff energy of 70 Ry, with Troullier-Martins norm-conserving pseudopotentials as implemented in \textsc{abinit} \cite{Gonze2009}.
The LDA energy levels were corrected to the first order in the subsequent $G_0W_0$ calculations, where the screened Coulomb interaction $W$ is expressed in random phase approximation (RPA) ($W=\varepsilon^{-1}v$), with $\varepsilon$ and $v$ being the dielectric function and the bare Coulomb interaction.
The inverse dielectric function $\varepsilon^{-1}$ was evaluated in the plasmon-pole approximation \cite{Godby1989}.
Spurious Coulomb interactions arising from the periodic images were truncated by a box-like cutoff \cite{Rozzi2006}.
The excitonic effect is finally taken into account by solving the BSE using the quasiparticle energies from the $G_0W_0$ calculations, which is equivalent to finding the eigenvalues of an effective two-particle excitonic Hamiltonian $H^\text{exc}$ as
\begin{equation}
H^\text{exc} =
\begin{pmatrix}
H^\text{res} & H^\text{cpl} \\
-[H^\text{cpl}]^\ast & -[H^\text{res}]^\ast 
\end{pmatrix}.
\end{equation}
In terms of the one-particle occupied and empty quasiparticle states ($v,c$) and quasiparticle energies $\varepsilon_i^\text{QP}$, the Hermitian resonant term ${H}^\text{res}$, containing the transitions from the occupied to the empty states ($v\rightarrow c$) of positive energies, can be expressed as $(\varepsilon_c^\text{QP} -\varepsilon_v^\text{QP}) \delta_{vv'} \delta_{cc'} + (K^d+2K^x)_{(vc)(v'c')}$, where $K^d$ and $K^x$ refer to the (screened) direct \textit{e-h} Coulomb interaction and the bare \textit{e-h} exchange interaction, respectively.
The direct interaction is calculated with the static screening in this work.
Analogously, the anti-resonant part $-[H^\text{res}]^\ast$ contains the $c\rightarrow v$ transitions of negative energies. 
The off-diagonal term $H^\text{cpl}=(K^d+2K^x)_{(vc)(c'v')}$ describes the coupling between the resonant and anti-resonant transitions.
Within the Tamm-Dancoff approximation (TDA), this coupling part is often neglected for optical absorption calculations since it is usually much smaller than the resonant part \cite{Onida2002}.
The TDA significant reduces the computational cost and becomes the routine for the BSE calculations for the excitations in various systems.
However, it is recently found that the overestimation of the excitation energy by TDA can be significant for organic molecules and nanostructures with confined excitations \cite{Ma2009,Palummo2009,Gruning2009}.
With an efficient iterative approach to the non-Hermitian $H^\text{exc}$ \cite{Gruning2009}, it is thus timely and necessary to go beyond the TDA in the present work.
The $G_0W_0$ and BSE calculations are carried out using the \textsc{yambo} code \cite{Marini2009}.

\begin{figure}
\includegraphics{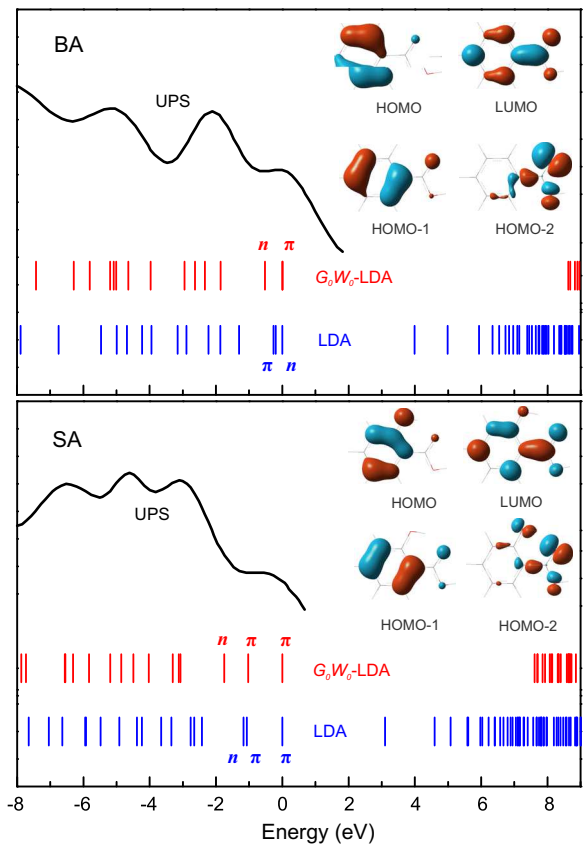}
\caption{\label{mol_mo} Molecular orbital energies of BA and SA calculated with LDA and $G_0W_0$-LDA, compared with experimental UPS data taken from Ref. \onlinecite{Tegenkamp2002}. The highest occupied orbitals are aligned to zero eV. The wavefunctions of the frontier orbitals by the PW1PW hybrid functional are presented.}
\end{figure}

We first consider the gas-phase molecules.
BA is the simplest aromatic carboxylic acid, with one carboxylic group attached to an aromatic ring.
BA can be functionalized to SA when one phenolic group is attached to the ortho position.
We show in Fig.~\ref{mol_mo} the molecular orbital energies calculated by LDA and $G_0W_0$-LDA.
The molecules are confined in a box with the dimension of $30.0 \times 30.0 \times 24.0$ Bohr$^3$. 
The $\mathbf{k}$-point is sampled at the $\Gamma$ point. 
Over 450 empty states are included for the response function (with a 4 Ry cutoff) and the Green's function for the self-energy (with a 68 Ry cutoff for the exchange self-energy). 
It is not surprising that LDA distorts the occupied molecular orbitals when compared to the experimental ultraviolet photoemission spectra (UPS) \cite{Tegenkamp2002}.
Notably, LDA gives a different order of the three highest occupied orbitals of BA with respect to the $G_0W_0$ result.
The $G_0W_0$ predicts that both the HOMO and HOMO-1 are delocalized $\pi$ states of the cyclic benzene ring, whereas it turns out that the LDA HOMO is the non-bonding state localized around the carboxylic group, which corresponds to the HOMO-2 in the $G_0W_0$ calculation.
This is interpreted as a consequence of the spurious self-interaction in the local and semilocal approximations of KS-DFT, which tends to shift the localized electronic states to higher energies \cite{Kummel2008}.
The effect of self-interaction is also observed for SA, where the LDA HOMO-2 is blue-shifted towards the HOMO-1 by about 0.6 eV compared to the quasiparticle energies.
We note that the self-interaction error can be effectively lifted by the nonlocal exact-exchange in hybrid functionals, and our PW1PW hybrid functional calculations \cite{Chen2009} are in line with the $G_0W_0$ quasiparticle energies for the occupied orbitals.
Being a dynamic theory, the $G_0W_0$ results are in quantitative agreement with the UPS experiment \cite{Tegenkamp2002}.
In particular, the $G_0W_0$ HOMO-3 levels for both molecules align well with the UPS peaks.

Moving to the unoccupied orbitals, we find that the quasiparticle energy corrections are rather substantial.
An interesting observation is that the quasiparticle correction to the LUMO+2 is about 2 eV smaller than that to the LUMO and LUMO+1 in the present calculations.
Instead of the $\pi^\ast$ anti-bonding character of the LUMO and LUMO+1, the wavefunction of the LUMO+2 is largely delocalized away from the molecule, a sign of a Rydberg orbital \footnote{As a result, the quasiparticle energy of the LDA LUMO is 0.2 eV higher than that of the LDA LUMO+2 for BA.}.
The Rydberg orbital is in general not correctly described by LDA because of the unrealistic exponential decay of the potential in the long-range.
In terms of the HOMO-LUMO gap, we see in Fig.~\ref{mol_mo} that the $G_0W_0$ quasiparticle energy gap is more than twice as large as the LDA gap.

\begin{table}
\caption{\label{S1} The lowest singlet excitation energies (in eV) of BA and SA calculated within BSE, in comparison with experimental values. The exciton binding energy $E_\text{B}$ is calculated as the difference between the $G_0W_0$ HOMO-LUMO gap ($E_\text{g}$) and BSE excitation energy.}
\begin{ruledtabular}
\begin{tabular}{lcccc}
$S_1$($\pi$-$\pi^\ast$) & BSE & Expt. & $E_\text{g}^{G_0W_0}$ & $E_\text{B}$ \\
\hline
BA & 4.67 & 4.56\footnotemark[1] & 8.82 & 4.15 \\
SA & 3.87 & 3.99\footnotemark[2] & 7.61 & 3.74 \\
\end{tabular}
\end{ruledtabular}
\footnotetext[1]{Reference \onlinecite{Ungnade1952}.}
\footnotetext[2]{Reference \onlinecite{Ernst1963}.}
\end{table}

\begin{figure}
\includegraphics{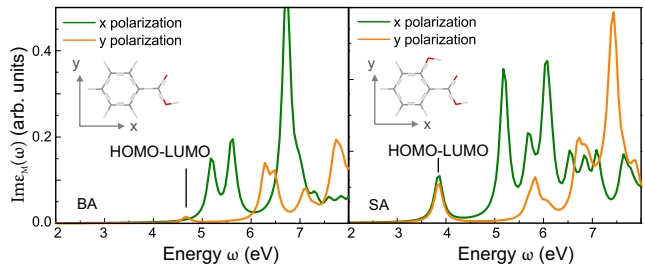}
\caption{\label{mol_bse} Optical absorption spectra of BA and SA obtained by full BSE calculations. The spectra are broadened by an artificial Lorentzian of 0.1 eV.}
\end{figure}

Referring to the lowest absorption peaks from available UV/visible spectra \cite{Ungnade1952,Ernst1963}, one notices that the $G_0W_0$ gaps, as the onsets in the $GW$-RPA independent-particle spectra, are also much larger than the experimental excitation energies (cf. Table~\ref{S1}), implying that the bound \textit{e}-\textit{h} interaction must play an important role in the optical absorption if the $G_0W_0$ gaps are reasonable.
When the \textit{e}-\textit{h} interaction is taken into account in the BSE, excellent agreement with experiment can be achieved in terms of the lowest excitation energy.
In the BSE calculations for the molecules, the cutoff energies for the direct ($K^d$) and exchange ($K^x$) \textit{e}-\textit{h} interactions are 2 Ry and 68 Ry, respectively. 
We include \textit{e}-\textit{h} pairs involving the transitions in an energy window of 33.2 eV in the BSE kernel. as shown in Table~\ref{S1}. 
It is seen in Table~\ref{S1} that the deviation between the BSE excitation energy and experimental optical peak is within 0.12 eV.
The coupling part in the excitonic Hamiltonian has a small effect on the lowest singlet excited state for the individual molecules, as the peak shifts by merely 0.1 - 0.2 eV using the TDA. 
The optical absorption in Fig.~\ref{mol_bse} is given by the imaginary part of the macroscopic dielectric function $\varepsilon_\text{M}(\mathbf{q}=0,\omega)$, where $\mathbf{q}$ is the momentum transfer. 
The lowest singlet excited state for both BA and SA largely correspond to the $G_0W_0$ HOMO $\rightarrow$ LUMO transition, \textit{i.e.} the $\pi \rightarrow \pi^\ast$ transition.
The polarization direction dependence of the lowest excitation peak can be understood by the symmetries of the HOMO and LUMO wavefunctions illustrated in Fig.~\ref{mol_mo}.
The \textit{e}-\textit{h} pair is very strongly bound, as the molecular exciton binding energies shown in Table~\ref{S1} are about half of the fundamental gaps.
Due to the nature of the $\pi \rightarrow \pi^\ast$ transition, the exciton is spatially localized at the aromatic benzene ring.
The strong excitonic effect is an important feature of the benzoic and its phenolic derivatives.

\begin{table}
\caption{\label{pol}Calculated $\pi-\pi^\ast$ gaps (in eV) of BA and SA on NaCl(001) by LDA and $G_0W_0$-LDA. The isolated molecule refers to the molecule detached from the surface while its geometry is kept fixed as that of the adsorbate. The molecular energy gap change upon adsorption is denoted by $\Delta E_\text{g}$. }
\centering
\begin{ruledtabular}
\begin{tabular}{l rr rr}
 & \multicolumn{2}{c}{BA} & \multicolumn{2}{c}{SA} \\
\cline{2-3}
\cline{4-5}
& LDA & $G_0W_0$ & LDA & $G_0W_0$\\
\hline
Adsorbate & 4.18 & 8.15 & 3.09 & 6.59 \\
Isolated & 4.21 & 9.11 & 3.08 & 7.66 \\
$\Delta E_\text{g}$ & $-0.03$ & $-0.96$ & $+0.01$ & $-1.07$ \\ 
\end{tabular}
\end{ruledtabular}
\end{table}

After examining the individual molecules, we now proceed with the adsorption on the NaCl(001) surface.
The molecules are placed on a four-layer (2$\times$2) NaCl(001) supercell, with a vacuum thickness of at least 15 \AA.
The LDA bulk lattice parameter ($a=5.64$ \AA) is used. 
We use a $\Gamma$ centered 3$\times$3 Monkhorst-Pack $\mathbf{k}$-point mesh (5 $\mathbf{k}$-point in the irreducible Brillouin zone).
The adsorption geometry optimized by the LDA functional is justified since the long-range van der Waals (vdW) interaction has a negligible influence on the adsorption configuration \cite{Chen2010}.
The molecules are weakly bound to the surface, with a binding energy of about 0.4 eV from the ionic contribution and 0.5 eV from the vdW interaction \cite{Chen2010}.
While the unit cell is small, the lateral intermolecular interaction is still limited, as is manifested by the small energy level dispersion of the HOMO and LUMO with respect to the $\mathbf{k}$ points.
For $G_0W_0$ calculations, we applied a total number of 960 bands in evaluation of the self-energy, and a cutoff of 60 Ry and 4 Ry for the exchange self-energy and the response function, respectively.
Since the lowest excited state of the molecules involves the $\pi \rightarrow \pi^\ast$ transition, the evolution of the $\pi-\pi^\ast$ energy gap upon adsorption on NaCl(001) was investigated and is presented in Table~\ref{pol} for BA and SA.
Note that in LDA the $\pi-\pi^\ast$ gap refers to the (HOMO-1)-LUMO instead of the HOMO-LUMO gap for BA.
Analogous to the gas-phase molecules, the $G_0W_0$ energy gaps of the adsorbates are significantly larger than the LDA ones.
In particular, we find that LDA predicts that the HOMO of BA is below the surface valence band maximum (VBM), which is clearly at variance with our earlier hybrid functional results \cite{Chen2009} as well as with the present $G_0W_0$ calculation.
As the non-bonding state localized at the carboxylic group is in resonance with the surface VBM \cite{Chen2009}, the misalignment of the energy levels in the LDA calculation is clearly associated with the self-interaction error, which places the non-bonding state too high in energy.
It is further revealed in Table~\ref{pol} that, while LDA yields almost identical energy gaps for the isolated and the adsorbed molecules, the quasiparticle energy gaps are reduced by about 1 eV when the molecules are in contact with the surface according to the $G_0W_0$ calculations.
The renormalization of the molecular energy gap in $G_0W_0$ is essentially an effect of the polarization in the surface due to the added or removed electron in the adsorbate, which in turn acts back on the adsorbed molecule and shifts its electron affinity and ionization energy level positions \cite{Neaton2006,*Freysoldt2009,*Garcia-Lastra2009}.
For example, the ionization energy level (HOMO) moves up by 0.36 eV whereas the electron affinity level (LUMO) moves down by 0.71 eV when a SA molecule is adsorbed on NaCl(001).
The asymmetric shifts suggests that the renormalization in the present case does not strictly follow the classical image potential theory, and polarizations between the molecules are likely to take place with a full monolayer coverage.
Decomposing the self-energy into the bare exchange ($\Sigma_x$) and the correlation ($\Sigma_c$) components, we find that the shifts of the energy levels arise solely from the change in the $\Sigma_c$.
The polarization effect, thus being a dynamic correlation effect, cannot be captured by the KS eigenvalues in standard DFT approximations.
Our results show that the effect can be pronounced even on wide-gap insulators with small dielectric constant, such as NaCl.

\begin{figure}
\includegraphics{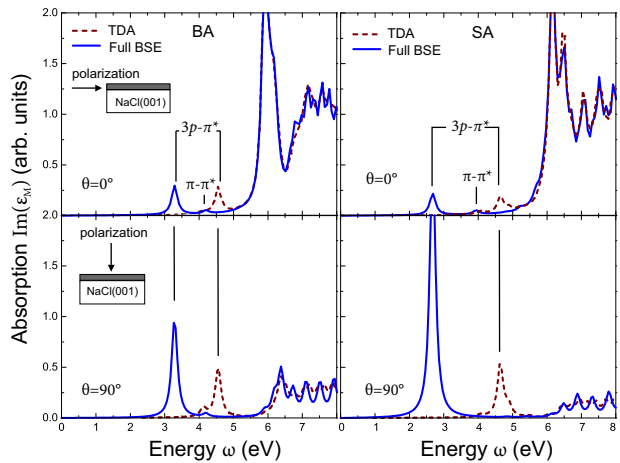}
\caption{\label{ads_exc}Absorption spectra of the BA and SA molecules adsorbed on NaCl(001) calculated within the Tamm-Dancoff approximation (TDA) and full excitonic Hamiltonian. The perturbative polarization direction with respect to the surface is denoted by $\theta$. The spectra are broadened by a 0.1 eV Lorentzian.}
\end{figure}

As the final step, the excitonic effect at the molecule-surface interface is addressed by BSE.
We include \textit{e}-\textit{h} pairs with energies up to 33 eV in the two-particle response function.
The cutoff energies for the $K^d$ and $K^x$ were chosen to be 3 Ry and 20 Ry, respectively. 
The lowest excitation energies are converged within 0.2 eV.
In Fig.~\ref{ads_exc} the optical spectra calculated within the TDA and the full BSE calculation are presented.
We first discuss the TDA results.
The first excitation peak appears at around 4.1 eV for both the BA and SA adsorbate systems.
This peak is assigned to the $\pi \rightarrow \pi^\ast$ transition of the molecule according to a full diagonalization of the TDA Hamiltonian.
Going from the parallel perturbing field ($\theta=0^{\circ}$) to the perpendicular polarization ($\theta=90^{\circ}$), the intensity of the $\pi \rightarrow \pi^\ast$ peak does not change much because the molecules are tilted on the surface.
Moving to higher photon energies, we see that the second lowest peak appears at 4.6-4.7 eV for both molecules, which primarily comprises the excitation from the surface VBM to the molecular $\pi^\ast$ orbital, \textit{i.e.} the Cl $3p \rightarrow \pi^\ast$ transition.
This peak is thus clearly attributed to a charge transfer exciton, which in principle arises from the direct overlap between the surface Cl states and the delocalized $\pi^\ast$ orbital of the adsorbate.
The observed charge transfer exciton is rather surprising, because in the ground state the charge transfer is very small from the surface to the molecule \cite{Chen2009}.
At 6.0-6.1 eV, the strongest absorption peak appearing in the parallel polarization stems from the NaCl(001) surface exciton, which lies about 1 eV below the bulk exciton due to the image states outside the surface \cite{Rohlfing2003}.
The peak related to the surface exciton is suppressed for the transverse field. 

When the coupling part is switched on in the full BSE calculation, the valence band $3p \rightarrow \pi^\ast$ peak experiences a sizable red-shift by 1.3-2.0 eV.
Remarkably, not only the excitation energy shifts, but the intensity of the $3p \rightarrow \pi^\ast$ transition also grows much stronger when $\theta=90^{\circ}$.
Such a large perpendicular transition dipole moment further evidences that the exciton is spatially distributed along surface normal.
The much stronger oscillator strength of this peak for the adsorbed SA molecule is benefited from its adsorption configuration, as the SA molecule is more parallel and closer to the surface than the BA \cite{Chen2010}, giving rise to a stronger overlap between the VBM and the $\pi^\ast$ orbital.
On the other hand, the lowest singlet molecular excited state and the surface excited state are not much influenced by the coupling part.
As a result, the lowest excited state for the adsorbate system is now replaced by the $3p \rightarrow \pi^\ast$ transition, with excitation energies of 3.3 eV and 2.7 eV for the BA and SA adsorbate systems, respectively.
Our results shows that the presence of the charge transfer exciton considerably lowers the effective excited energy of the molecule-insulator system, which may serve as the key ingredient in the contact charging model.

The striking failure of TDA reveals that the coupling part becomes comparable to the resonant part in the excitonic Hamiltonian for the $3p \rightarrow \pi^\ast$ transition.
In fact, we find that even without the direct interaction $K^d$, the coupling term with the exchange interaction $K^x$ alone already shifts the lowest excitation energy from 4.6 eV in TDA to 3.0 eV in the case of SA.  
Therefore, the $3p \rightarrow \pi^\ast$ exciton has a prominent exchange interaction between the electrons and holes, making TDA insufficient in describing this excited state.

To conclude, we find that even when a conjugated organic molecule is weakly adsorbed on an wide-gap insulating surface, the presence of the exciton at the molecule-insulator interface can provide a much more effective channel for electronic excitations than the intramolecular transitions.
The formation of the lowest excited state involving transitions from the surface to the molecule is driven by the sizable \textit{e}-\textit{h} exchange interactions.

\begin{acknowledgments} 
This work was supported by K+S AG. We acknowledge generous computation time allocated at HLRN. 
\end{acknowledgments}

\bibliography{excitation}

\end{document}